\documentstyle[aps]{revtex}
\def\be {\begin{equation}}
\def\ee {\end{equation}}
\def\ba {\begin{eqnarray}}
\def\ea {\end{eqnarray}}

\def\a  {\alpha}

\def\d  {\delta}

\def\o  {\omega}

\def\p  {\pi}
\def\P  {\Pi}

\def\le {\left}
\def\ri {\right}

\begin{document}
\draft
\title{Quantum Mechanics of Charged Black Holes}
\author{Andrei Barvinsky $^a$,
Saurya Das $^b$
and Gabor Kunstatter $^b$}
\address{$^a$ Lebedev Physics Institute and Lebedev Research Center
in Physics,
Leninsky Prospect 53, Moscow 117924, RUSSIA}
\address{$^b$ Dept. of Physics and Winnipeg Institute for
Theoretical Physics,
University of Winnipeg, Winnipeg, MB R3B 2E9, CANADA \\
(Email: barvin@td.lpi.ru, saurya@theory.uwinnipeg.ca,
gabor@theory.uwinnipeg.ca)}
\maketitle

\begin{abstract}
We quantize the spherically symmetric sector of generic  charged black
holes. Thermal properties are encorporated by imposing periodicity in
Euclidean time, with period equal to the inverse Hawking temperature of
the black hole. This leads to an exact quantization of the area (A) and 
charge (Q) operators. For the Reissner-Nordstr\"om black hole, $A=4\pi G \hbar 
(2n+p+1)$
and $Q=me$, for integers $n,p,m$. Consistency requires the fine structure 
constant to
be quantized:  $e^2/\hbar=p/m^2 $. Remarkably, vacuum fluctuations exclude 
extremal black holes
from the spectrum, while near extremal black holes are highly quantum
objects. We also prove that horizon area is an adiabatic invariant.
\end{abstract}


Bekenstein and Hawking\cite{BH} showed  almost thirty years ago
that black holes possess intriguing  thermodynamic properties
which  make them rich theoretical  laboratories for testing
theories of quantum gravity. Although candidate theories for
quantum gravity exist (such as string theory and quantum
geometry),  the microscopic origin of thermodynamic behaviour is
still largely the subject of conjecture. It is therefore
important to learn as much as possible about the quantum
behaviour of black holes without assuming a specific underlying
microscopic theory.  One very natural question that arises in
this context concerns the quantum mechanical spectrum of the
observables associated with charged black holes. Based on the
conjecture that horizon area is an adiabatic invariant, as well as from
other considerations, it was postulated that for neutral
black holes, the area spectrum is discrete and uniformly spaced;
i.e. $ A \propto n ~~$, where $n$ is an integer \cite{bm0}.

In this letter, we follow an argument originally presented in
\cite{bk} to quantize the spherically
symmetric sector of generic charged black holes. 
Our starting point is the assumption that it is possible 
to encorporate the thermodynamic behaviour of  black holes into
a quantum description by
imposing periodicity in Euclidean time, with period equal to the inverse
Hawking temperature of the black hole. This single assumption 
 allows us to (i) derive an exact quantized area
spectrum; 
(ii) derive the spectrum of electric charge and (iii) show that
black hole quantization places stringent restrictions on the fine
structure constant; 
(iv) prove that horizon area is an adiabatic invariant.
We emphasize that our  analysis is quite
general and is not tied to a specific model or theory of
gravity, in contrast with other derivations of qualitatively
similar spectra \cite{others}. It applies, for example,  to
charged as well as neutral black holes in Einstein-Maxwell theory
in any dimension, and in fact, even to the $3$-dimensional
rotating BTZ black hole (where the angular momentum plays the
role of the electric charge).

Note that  our goal here is  not to explain the microscopic source of the 
thermodynamic behaviour. We simply encode it into the boundary conditions
and observe the consequences.  The formalism of Euclidean
quantum field theory, as is well known, can originate from two
distinctively different physical situations -- from the
description of thermodynamical ensemble (statistical, i.e. not
pure, state) or from the description of classically forbidden
transitions between pure states -- quantum mechanical underbarrier
tunneling. Quite amazingly, in quantum gravity these two functions
of the Euclidean formalism are not clearly separated.
 Indeed, the Euclidean section of the
Schwarzschild solution can, on one hand, be regarded as a saddle
point of the path integral for the statistical partition function
and, on the other hand, can be viewed as a classical configuration
interpolating in the imaginary time between the two causally
disconnected spacetime domains:the right and left wedges of the
Kruskal diagram. Our requirement of periodicity in  imaginary
time  can be viewed as a kind of consistency of quantum
states in these two domains, or the finiteness of the
semiclassical underbarrier transition amplitude between them
(remember that the Hawking periodicity requirement is based on the
absence of conical singularity which is, in its turn, motivated by
the regularity of the semiclassical distribution). So
amplitudes not satisfying this periodicity requirement can be regarded as
suppressed.

We restrict consideration to black hole spacetimes
that are static and can be parametrized by only
two coordinate invariant parameters, which we choose to be the
mass $M$ and charge $Q$. This basically assumes a Birkhoff-like
theorem, and forbids the presence of monopole gravitational or
electromagnetic radiation.  With this assumption, there exists a
coordinate system in which the metric takes form:
    \be
    ds^2=-f(x;M,Q)dt^2 + {dx^2\over f(x;M,Q)}
    + r^2(x)~d\Omega^{(d-2)}\, .
    \label{metric2}
    \ee
where $x$ is a radial coordinate. The function
$f(x;M,Q)$ is uniqely determined by the
requirement that $-g_{tt}$ and $g_{xx}$ are inverse proportional to one
another, and the 
location of the
horizon $x_h=x_h(M,Q)$ is given implicitly by: $f(x_h;M,Q)=0$.

An elegant way to extract thermodynamic information about black
hole spacetimes is to euclideanize the solution and concentrate
on the  `near-horizon' region. By a suitable choice of Euclidean
time $t_E= -it$ and spatial coordinate $R(x)=
\sqrt{{f(x;M,Q)}/\le( f'(x_h;M,Q) \ri)}$, one can put the metric
near the horizon $(R \rightarrow 0)$ into the form, $ ds_E^2 =
dR^2 + R^2 d\a^2$, where $\a:= t_E f'(x_h;M,Q/)/2$. To avoid a
conical singularity at the horizon, $\a$ must have the period
$2\p$, implying that $t_E$ must be periodic with range $0\leq t_E
\leq 4\pi/f'(x_h;M,Q) \, .\label{period}~ $ It follows from
finite temperature quantum field theory that the (Hawking)
temperature associated with this black hole is the inverse of the
Euclidean period, i.e. : $~T_H(M,Q)= {\hbar f'(x_h;M,Q)/ 4\pi}~.$
This will play a key role in what follows. The Bekenstein-Hawking entropy,
$S_{BH}(M,Q)$, 
of the black hole is defined generically  by requiring it to obey the first law of
thermodynamics:
 \be~\delta M = T_H(M,Q)
\delta S_{BH}(M,Q) + \Phi(M,Q)\delta Q ~,~~ \ee where $\Phi(M,Q)$
is the electrostatic potential at the horizon. Given $T_H$ and the
electrostatic potential, this determines $S_{BH}(M,Q)$ up to an additive constant, which is fixed by requiring the Bekenstein-Hawking entropy to vanish when
the mass and charge both vanish. For spherically symmetric black holes in 
any dimension, this yields the usual relationship between
the entropy and the area of the outer horizon: $S_{BH}= A/4G\hbar$.
 For example, in the case of
 the Reissner-Nordstr\"om  black hole, $
T_H = 2\hbar {\sqrt{M^2 - Q^2}}/A $, $A=4\pi r^2_+$ and  $\Phi= Q/r_+$,
where $r_+=(GM+\sqrt{G^2M^2-GQ^2})$.

Since $M$ and $Q$ are assumed to be the only diffeomorphism
invariant parameters, the reduced action governing the dynamics
of the  spherically symmetric sector of isolated, generic charged
black holes in any theory must be of the form \cite{dk,mk}:
    \be
    I^{red}= \int dt \left(P_M \dot{M} +
    P_Q \dot{Q} - H(M,Q)\right) \, ,       \label{reduced action}
    \ee
where $P_M$ and $P_Q$ are the conjugates to $M$ and $Q$,
respectively. The exact expression for the Hamiltonian is
irrelevant: the fact that it is independent of $P_M$ and $P_Q$
alone ensures that $M$ and $Q$ are constants of motion. Of
course, one can also arrive at this reduced action via a rigorous
Hamiltonian analysis of the  spherically symmetric charged black
hole spacetimes.  For details, see \cite{dk,mk}.

For boundary conditions which preserve the so called
Schwarzschild form (\ref{metric2}) of the metric at either end of
a spatial slice, $P_M$ can be shown to be proportional to the
difference between the Schwarzschild times at either end of the
slice \cite{kuchar,gkl,thiemann}. Moreover, the momentum $P_Q$ is
related to $P_M$ by means of the following relation : \be \delta
P_Q =- \Phi \delta P_M + \delta \lambda \, ,\label{momentum
relation} \ee where $\d P_Q$ and $\d P_M$ refer to variations
under a change in boundary conditions and $\delta \lambda$ is the
variation in $U(1)$ gauge transformation $\lambda$ at the
horizon.

In order to quantize we need to know the boundary conditions on the
phase space variables. We require $M>0$, $T_H(M,Q) \geq 0$ and $Q$ to be
real. Using the expressions derived in \cite{dk}, it can be shown that positivity of the Hawking temperature leads generically to a condition of the form:
    \be
    S_{BH}(M,Q)\geq S_0(Q) ~. \label{entropy minimum}
    \ee
where the equality is achieved in the limit of extremal black holes.
The lower bound $S_0(Q)$ on the Bekenstein-Hawking entropy is a uniquely determined function of $Q$ for each theory in the class under consideration. 
For example, for the Reissner-Nordstrom black hole, $S_0(Q)=\pi Q^2/\hbar$.
   
Until this point our anlaysis has been more or less standard.
We now go to the Euclidean sector where the  time difference
$P_M$ becomes imaginary as well as periodic, with period given by
$T_H^{-1}(M,Q)$. Although it is possible to derive a black hole spectrum
by imposing periodicity of the Lorentzian time coordinate\cite{others}, the motivation for
the periodicity is more problematic than in the Euclidean sector.  In the present case, the
procedure is well defined and consistent.  Essentially, we start
 with the reduced Hamiltonian and action as given  in the Lorentzian
sector, which is of precisely the same form as Eq.(3), and analytically 
continuing to Euclidean time before quantizing. As a direct consequence the
momenta conjugate to $M$ and $Q$ are pure imaginary. However, since the
Hamiltonian is independent of these conjugate momenta, it does not change
its form. Thus, Euclideanization  merely generates
an overall factor of $i$ in front of the reduced action and keeps the
dynamics unaltered. Ultimately, the physical relevance of our derived spectra
will rest on the connection between the charge and mass eigenstate wave
functions that we construct in the Euclidean sector using Hamiltonian
techniques, and their counterparts in the Euclidean path integral formulation
of quantum gravity.

Periodic boundary conditions on phase space variables are familiar in classical
mechanics. Akin to the action-angle formulation of the harmonic
oscillator, we can `unwrap' our gravitational  phase space, by
transforming to a set of unrestricted variables. Consider the
following transformation $(M,Q,P_M,P_Q) \rightarrow
(X,Q,\P_X,\P_Q)$, which directly encorporates the correct
periodicity of $P_M$:
    \ba
    &&X=\sqrt{\hbar B(M,Q)/\pi}
    \cos(2\pi P_M T_H(M,Q)/\hbar) \, ,\nonumber\\
    &&\P_X= \sqrt{\hbar B(M,Q)/\pi }
    \sin (2\pi P_M T_H(M,Q)/\hbar)    \, ,        \label{grav}\\
    &&Q=Q \, ,\nonumber\\
    &&\P_Q=\P_Q(M,P_M,Q,P_Q) \, ,  \nonumber
    \ea
where the functions $B(M,Q)$ and $\Pi_Q(M,P_M,Q,P_Q)$ will be
determined shortly. Direct calculation shows that this
transformation is canonical if and only if:
    \be
    {\partial B\over\partial M}
    ={1\over T_H(M,Q) } ~~~,~~~~~
    P_Q =\Pi_Q+
    P_M T_H {\partial B\over\partial Q}\, .  \label{canonical 1}
    \ee
From the first law of black hole mechanics we know that $
\partial S_{BH}/\partial M = T_H^{-1}(M,Q)$. Thus we conclude:
    \be
    B(M,Q)= S_{BH}(M,Q)+F(Q) \, ,\label{B1}
    \ee
where $F(Q)$ is an arbitrary function of the charge. Combining
(\ref{B1}) and (\ref{grav}) we get:
    \be
 S_{BH}(M,Q) + F(Q) = {2\pi\over\hbar} \left({1\over 2}X^2
    + {1\over 2}\P_X^2\right), \label{B2}
    \ee
which shows that the subspace $(X,\P_X)$ has a `hole' of radius
$[ S_0(Q) + F(Q)]^{1/2}$, the interior of which is inaccessible.
To remove potential quantization ambiguities, we choose $F(Q) = -
S_0(Q)$, thus removing the perforation and rendering the phase
space complete. As a bonus, this
automatically ensures that the inequality (\ref{entropy minimum})
is satisfied in a natural way. The
extremal limit now gets mapped to the origin of the new phase
space. With this choice, $\Pi_Q$ is uniquely determined to be:
    \be
    \Pi_Q = {\hbar\over e}\chi
    + {\hbar\over 2\pi}S'_0(Q)\alpha, \label{canonical 2}
    \ee
where $'=d/dQ$ , $\chi = (e/\hbar)( P_Q+\Phi P_M)$ and $\alpha =
2\pi P_M T_H(M,Q)/\hbar$.

From (\ref{B2}) it follows that the operator $S_{BH}-S_0(Q)$ is
precisely the Hamiltonian of a simple harmonic oscillator with
the mass and frequency both equal to unity. Since $-\infty \leq
X, \P_X \leq \infty$, standard quantization yields the spectrum:
    \be
    S_{BH}= 2\pi \left(n+{1\over 2}\right)+S_0(Q)
    \qquad n = 0,1,2,... \label{entropy spectrum}
    \ee
where we have assumed the usual harmonic oscillator factor ordering
for the operators $X$ and $P_X$ in constructing the quantum version of
(\ref{B2}).
A remarkable feature of (\ref{entropy spectrum})
 above is that vacuum fluctuations
exclude extremal black holes $(S_{BH}=S_0)$ from the quantum
spectrum. Another important result,one that is independent of the choice
of factor ordering, is that near-extremal states are highly
quantum-mechanical objects ($n \sim 0$), even for large values of
$M$ and $Q$.

To quantize the electromagnetic sector, we note from
(\ref{momentum relation}) that for compact gauge group $U(1)$,
$\chi:= e\lambda/\hbar=e( P_Q+ \Phi P_M)/\hbar$ has period
$2\pi$, where $e$ is the electromagnetic coupling. Thus from
(\ref{canonical 2}), $\Pi_Q$ is a function of two angular
coordinates $\chi$ and $\alpha$ which, according to arguments
given above are both periodic with period $2\pi$. We must
therefore identify the phase space points
    \be
    (Q, \Pi_Q) \sim
    (Q, \Pi_Q + 2\pi n_1 {\hbar / e}
    +  n_2 \hbar S'_0(Q)). ~~                   \label{identify}
    \ee
for arbitrary integers $n_1$ and $n_2$.
In the coordinate representation, ${\hat Q}=-i\hbar
\partial / \partial \Pi_Q$, the wave functions for
charge eigenstates take the form $\psi_Q(\Pi_Q) = {\rm
(const)}\times \exp(i Q \Pi_Q/\hbar)$. The spectrum of $Q$ is
restricted by the requirement that the wave function be 
single valued under the identification (\ref{identify}): for each 
admissable  $Q$,  for all integers $n_1$ and $n_2$ there must exist
a third integer $n_3$ such that:
\be
n_1 {Q/ e} + n_2 {Q S_0'(Q)/ 2\pi} = n_3 .
\label{constraint}
\ee
This in turn requires that
    \be
    {Q\over e}=m~~,~~~
    {Q\over 2\pi} S_0'(Q)=p,              \label{e-spectrum}
    \ee
for integer $m$, $p$. To see this,  suppose that there
exists some values of $n_1$, $n_2$ and $Q$ for which (14) holds for some 
$n_3$. If we increase the value of $n_1$ by 1, the value of $n_3$ increases
by $Q/e$, so it is necessary that $Q/e=m$, for some integer $m$ in order that
the shifted $n_3$ be an integer. Similarly, if we increase $n_2$ by
one, the second relation in (15) emerges as necessary.

 The first of the conditions (\ref{e-spectrum})
 gives the expected result
that  $Q$ must be an integer multiple of the fundamental charge
$e$. However, the second condition can only be satisfied if $e$
satisfies a subsidiary, and totally unexpected condition. The
specific form of this condition depends on the theory under
consideration. For Reissner-Nordstr\"om black holes in four
dimensions, $S_0(Q)= \pi Q^2/\hbar$ and (\ref{entropy spectrum})
and the second of equations (\ref{e-spectrum}) translate to:
    \be
    S_{BH}=2\pi n+\pi(p+1),~~~~
    Q^2=p~\hbar .              \label{charge1}
    \ee
The integer $p$ determines the charge of the quantum black holes
and hence its  minimum entropy $S_0=\pi(p+1)$, whereas $n$
determines the excited level of the black hole over the
``vacuum'', $n=0$. Finally, the first of equations
(\ref{e-spectrum}) requires
    \be
    {e^2\over\hbar}={p\over m^2}   \, .   \label{fsqc}
    \ee
Thus, the fine structure constant $e^2/\hbar$ must be a rational
number. For the d-dimensional generalization of these results see
\cite{bks}.

For one dimensional periodic systems, it is well known that
the integral $ {\cal J}_X= \oint \P_X~dX$ is an adiabatic
invariant. Thus, in the present case, if we treat $Q$ as a slowly 
varying parameter, it follows that 
${\cal J}_X  =  \pi ({A-4G~\hbar S_0(Q)})/{4G}$ is an adiabatic invariant. 
Consequently, away from extremality
($A >> 4G\hbar S_0(Q)$), this is consistent with Bekenstein's 
conjecture \cite{bm0} that the horizon area of a charged black hole is
an adiabatic invariant. 

The  expression (\ref{entropy spectrum}) has fascinating
consequences. First of all, it implies that extremal black holes,
for which $S_{BH}=S_0(Q)$, are not in the physical spectrum.
Secondly, if we interpret the entropy in terms of statistical
mechanics (\ref{entropy spectrum}) tells us that the degeneracy
of the $n$-th level is: $ g(n) = \exp[2\pi (n+1/2) + S_0(Q)]$.
Thus, the ground state is degenerate ($g(0) \neq 1$). It is
tempting to conjecture that this Planck size remnant provides
clues about the information loss problem associated with the
endpoint of Hawking radiation. Finally, (\ref{entropy spectrum})
allows Hawking radiation to be emitted when a black hole jumps
from one quantum entropy level to another. For a
Reissner-Nordstr\"om black hole, the fundamental frequency of
emission of a neutral quantum, $\o_0$ satisfies: $ S(M+ \hbar
\o_0,Q) - S(M,Q) = S(n+1) - S(n) = \p~,~ $ from which it follows
that $ \o_0 = (r_+ - r_-) \p/A$. In the Schwarzschild limit $Q
\rightarrow 0$, $\o_0 \sim 1/M$, agreeing with that found in
\cite{bm0}. Since the mean frequency of the Planck distribution
of Hawking radiation lies at $T_H \sim 1/M$, the radiation
spectrum consists of widely separated spectral lines, and
deviates considerably from the continuum originally predicted by
Hawking, no matter what the temperature. This turns out to be 
a generic feature of our spectra, valid for all black holes, charged
or uncharged\cite{bks}.

To summarize, we have looked at quantum gravity in the
spherically symmetric, charged black hole sector. To encorporate
the thermodynamic information, we assumed periodicity of the
momentum conjugate to the black hole mass. 
It is particularly important to stress that we used only 
very general features of black hole dynamics and thermodynamics. Consequently,
despite possible factor ordering ambiguities in our analysis, 
the following predictions are expected to
 be valid at least at the semi-classical level:
1)black hole area is an adiabatic invariant, hence its
quantum spectrum is equally spaced, 2) near
extremal black holes are highly quantum objects, and 3) the radiation 
spectrum of black holes is discrete, irrespective of the temperature.
Finally, 
 (\ref{entropy spectrum})
 and (\ref{e-spectrum}) imply that black holes emit and absorb
quanta whose charges are multiples of $e$, which itself is not
arbitrary, but quantized in terms of integers $m$ and $p$. Thus,
in analogy with the Dirac charge quantization condition in  the
presence of a magnetic monopole, the presence of charged black
holes puts constraints on the fine structure constant. This is
also reminiscent of the `big-fix mechanism' advocated by
Coleman,   wherein the fundamental constants of nature are
supposed to be fixed by the presence of wormholes and baby
universes \cite{bigfix}. Although, a priori, it is not clear  how
the experimentally measured value of the fine structure constant $4\p
\hbar/e^2 = 137.03608...$ can be reproduced accurately as the
ratio of integers that are not too large, as required by
(\ref{fsqc}), we believe that our results reveal some intriguing
features of the quantum mechanics of black holes and merit
further study.

While this paper was being completed, we became aware of two papers where the
spectra of charged black holes was investigated \cite{VW,MRLP}.
Although their results bear qualitative resemblance to ours, their
analysis and quantitative results differ considerably from ours.

\vspace{.5cm} \noindent {\bf Acknowledgments}

\noindent  S.D. and G.K. thank Viqar Husain for helpful comments
and encouragement. G.K. also thanks Valeri Frolov and the gravity
group at the Theoretical Physics Institute, University of Alberta
for useful discussions.  A.B. is grateful to V.Rubakov for
helpful discussion. We acknowledge the partial support of the
Natural Sciences and Engineering Research Council of Canada. This
work was also supported by the Russian Foundation for Basic
Research under the grant No 99-02-16122, the grant of
support of leading scientific schools No 00-15-96699 and
 in part by the Russian Research program
``Cosmomicrophysics''.

\end{document}